# Discovery of a 1D edge mode in a magnetic topological semimetal

Avior Almoalem[1], Rebecca Chan[1,2], Brinda Kuthanazhi[3,4], Juan Scmidt[3,4], Jose A. Moreno[3,5], Hermann Suderow[5], Paul Canfield[3,4], Taylor L. hughes[1,2] & Vidya Madhavan[1]

1. Department of Physics and Materials Research Laboratory, Grainger College of Engineering, University of Illinois at Urbana-Champaign, Urbana, IL, USA
2. Anthony J. Leggett Institute for Condensed Matter Theory, University of Illinois, Urbana, IL, USA
3. Ames Laboratory, Ames, Iowa, USA
4. Department of Physics and Astronomy, Iowa State University, Ames, IA, USA
5. Laboratorio de Bajas Temperaturas y Altos Campos Magnéticos, Unidad Asociada UAM-CSIC, Departamento de Fisica de la Materia Condensada, Instituto Nicolas Cabrera and IFIMAC, Universidad Autonoma de Madrid, E-28049 Madrid, Spain

**Abstract**

In rare-earth monopnictides like NdBi, the interplay between magnetism and topology results in an extremely unusual topological semimetal phase which simultaneously hosts Weyl points with Fermi arcs as well as massive and massless Dirac cones. A central question in this class of materials is whether ferromagnetic surfaces with gapped Dirac cones can also host robust well-defined chiral edge states. In this study, we use spin-polarized scanning tunneling microscopy (SP-STM) and spectroscopy to investigate the correlation between the magnetic and topological properties of NdBi. By combining SP-STM imaging with quasiparticle interference, we identify distinct signatures of both antiferromagnetic and ferromagnetic surface terminations and correlate them with their respective band structures. Crucially, we demonstrate that step edges on the ferromagnetic surface which serve as magnetic domain walls host well-defined one-dimensional (1D) edge modes that vanish above the Néel temperature. Our findings position NdBi as a promising platform for further explorations of 1D chiral edge modes and future realizations of Majorana states in proximitized rare-earth monopnictides.

## Introduction

A time-reversal invariant 3D topological insulator that develops antiferromagnetism (AFM) can remain topological, i.e., with a quantized magneto-electric polarizability and surface states, if the combined symmetries of time-reversal and translation are preserved [1]. Interestingly, theory shows that ferromagnetic surface terminations of the AFM order will open gaps in the surface Dirac cones and can host gapless chiral modes on magnetic domain walls [2-8]. These chiral domain-wall modes serve as a key manifestation of exotic quantum phenomena including axion electrodynamics and the quantum anomalous Hall effect, and they can generate chiral Majorana fermions when brought into proximity with a superconductor [9-23].

The rare-earth monopnictide NdBi presents an ideal platform for investigating these phenomena as the bulk is both topologically non-trivial and antiferromagnetic. However, this material does not strictly qualify as an antiferromagnetic topological insulator (AFTI) since the gapped topology coexists with some gapless bulk modes, including putative 3D Dirac cones. Since this system still hosts topological Dirac surface states, a critical question is whether magnetic domain walls will give rise to well-localized chiral modes even though the bulk is nominally gapless. Fortunately, there is reason to be optimistic since experiments on the Weyl semi-metal phase of $Co_3Sn_2S_2$ indicate that chiral modes can be well-localized on surface step edges, even when the bulk is nominally gapless [24].

We begin with a description of the known properties of NdBi. The paramagnetic phase of NdBi [25-27] hosts topologically non-trivial bands characterized by a Fu-Kane topological index of $(\nu_0, \nu_1, \nu_2, \nu_3) = (1,0,0,0)$ [25-27], which indicates a strong topological insulator (STI). Band inversion occurs between the Bi-6p and Nd-5d orbitals along the Γ-X line [26,27]. Below the Néel temperature of 24K, NdBi is an AFM with a rock-salt crystal structure [25,26]. Once the material transitions into an AFM phase upon cooling, time-reversal and translational symmetries are individually broken; however, their combination remains preserved in the bulk [1,15,16]. As shown in figure 1a, for a given crystal, both FM and AFM surface terminations are realized as different facets of the same bulk spin structure. In fact, within one single crystal, the material may host different bulk domains with the surface termination being either FM or AFM.

ARPES and DFT calculations of NdBi suggest that there is a strong correlation between the electronic structure and the surface magnetic order. Nano-ARPES studies reveal two types of areas: one where surface Dirac cones, electron like surface states, and Fermi arcs are all present, and another where only gapped Dirac cones exist. DFT calculations indicate that the Fermi arcs and electron like surface states are seen on surfaces where spins are AFM-aligned in-plane. The calculations also show that all these features should be absent on ferromagnetic (FM) surface terminations [25-27]. A ferromagnetic surface coupled with topologically non-trivial surface states offers a promising platform for the emergence of 1D chiral modes, which can be systematically explored and analyzed [1,25] as was proposed earlier in $MnBi_2Te_4$ [29].

To date, the relationship between distinct types of magnetic surfaces and the associated emergent electronic phenomena has not been experimentally established in NdBi. The lack of a comprehensive understanding of the sample's different electronic phenomena has posed a significant challenge to developing a complete theoretical framework for the material [25-28,30-

31]. In this work, we present an experimental study that seeks to conclusively establish these connections and investigates the existence of the chiral edge state.

**Results**

As previously discussed, NdBi is a bulk antiferromagnet in which Nd and Bi atoms adopt a cubic, rock-salt crystal structure. In the magnetically ordered phase, distinct surfaces having different magnetic textures emerge as shown in Fig.1a. To carry out STM studies, NdBi samples were cleaved at ~300K and at pressures lower than 5E-10 torr, before being directly inserted into the STM head held at 1.9K. To understand the magnetic structure and correlate it with the surface electronic structure, STM data were acquired both with spin-polarized Cr tips and un-polarized W-tips, at B=0T. Topography obtained with W-tips (Fig.1b) reveal surfaces consistent with imaging the non-magnetic NdBi unit cell where only one species (either Nd or Bi) is visible. However, images obtained with the Cr tip reveal two different surfaces, one with the same periodicity as the W-tip, and the other with a larger periodicity where the distances between the atoms is $\sqrt{2}$ larger (Fig. 1c).

**AFM Surface and QPI signature**

We first focus on surfaces that display a larger unit cell with spin-polarized tips. This domain comprises the majority of the sample's surface. In the tunneling process with a spin polarized tip, the matrix element is suppressed for spins anti-parallel to the tip apex-spin, and enhanced for parallel spins [32,33], (Supplementary Figure S1). Consequently, the tunnel current from surface atoms is either amplified or diminished according to their spin orientation. Thus, the images with the larger unit cell correspond to an antiferromagnetic configuration in which each spin is surrounded by anti-parallel neighbors (AFM surface). The spin structure can also be seen in the corresponding fast Fourier transform (FFT) of the AFM surfaces obtained with Cr tips, which reveal characteristic peaks at wavevector $q = 1/\sqrt{2}$ and a 45° rotation relative to the Bragg peaks seen in the FFT of the W tip images, as shown in the inset to Fig. 1d, e. The real-space and momentum-space periodicity is as expected from the AFM unit cell as shown in Fig. 1c. While the surfaces may look different with W- and Cr-tips, the dI/dV spectra show similar overall features: a dip near $E_F$ as well as peak-like features at similar energies ~ -10, 5 and 30 mV (Fig. 1d, e).

To elucidate the relationship between the magnetic structure and the emergent electronic phenomena, we investigate the electronic structure of the AFM surface via quasi-particle interference (QPI) measurements. In general, dI/dV maps provide insight into spatial variations of the local density of states (LDOS), which capture the interference patterns generated by the scattering of quasi 2D bulk states, surface states, and Fermi arcs [34–37]. By applying Fourier analysis to these maps, the contributing scattering processes can be decomposed according to their respective momentum transfers. A large-area topographic-image of an AFM surface featuring adatoms (bright spots) and vacancies (dark spots) is shown in Fig. 2a. Lattice defects, originating from the adatoms or vacancies, can induce quasi-particle scattering events. As shown in Fig. 2b, these events generate $C_2$-symmetric wave-like patterns. Consequently, the FFT of the dI/dV maps reveal wave vectors along the Γ–M direction (Fig. 2c).

The momentum-space location and $C_2$-symmetric structure of the QPI are reminiscent of the non-topological surface states and Fermi arcs measured by ARPES which exhibit $C_2$ symmetry [25-27]. DFT calculations [27] indicate these $C_2$ symmetric features are associated with the AFM surface. To confirm the origins of the measured QPI we use the ARPES-derived $C_2$ symmetric bands to simulate the expected QPI signal by calculating the join density of states (JDOS), using the auto-correlation of the Fermi surface, which we then compare with our QPI data. Since QPI signals are predominantly derived from two-dimensional (or quasi-two-dimensional) electronic states, we use the emergent electron- and hole-like surface states, as well as Fermi arcs, observed in a recent ARPES paper [25, 26] (Fig.2f) for the QPI simulation. We note that the topological Dirac surface states are not included here since backscattering is forbidden and they will not generate a QPI signal. A similar JDOS calculation which includes the Dirac cones is shown in supplementary Figure S2. All QPI data shown in the paper were symmetrized according to the intrinsic $C_2$ symmetry of the surface states and Fermi arcs. The corresponding unsymmetrized data is provided in Supplementary Figure S3.

Upon comparing the JDOS shown in Fig. 2d with the FFT in Fig. 2c, we conclude that the primary scattering processes observed include: (i) scattering from hole-like surface states around the Γ point (Q1), (ii) scattering between hole-like Fermi arcs (Q2), and (iii) scattering between electron-like surface states (Q3). The corresponding dispersions are shown in Fig. 2e and are consistent with ARPES data from previous studies [25-27]. The QPI signal weakens considerably for energies below –50 meV and vanishes entirely at –120 meV (Supplementary Figure S4), which is also consistent with the dispersion characteristics of surface states and Fermi arcs as observed in ARPES studies. Identical QPI patterns are observed using both the Cr and W tip, and the QPI remains consistently aligned along the Γ–M direction of the nonmagnetic Brillouin zone (Supplementary Figure S4). It is noteworthy that the QPI cannot be accurately reproduced without accounting for surface states close to the Γ point (Q1), shown in Fig. 2f where the Fermi level contour of the surface states, and the corresponding scattering vectors, are presented.

To further correlate the AFM phase with the emergent QPI, we conducted temperature-dependent measurements at 1.9 K and 25 K, i.e., below and above $T_{Néel}$. Both data sets, above and below the transition temperature, were acquired under the same STS conditions and at the exact same location. Within the magnetic phase, strong QPI signals and distinct AFM peaks in the FFT are observed (Supplementary Figure S5). Upon heating the sample slightly above $T_{Néel}$, the AFM peaks disappear as expected. Moreover, signatures of QPI are absent at all energies above $T_{Néel}$ which indicates that the associated electronic states responsible for the QPI are likewise absent. This is fully consistent with ARPES data which show that the Fermi arcs and electron-like surface states as presented in Fig. 2f emerge exclusively in the AFM phase [25,26].

**FM surface and 1D Edge Modes:**

We now move on to the surface with broken time reversal symmetry i.e., the ferromagnetic surface. As show in figure 1a, this surface represents the FM termination of stacked layers of spins in which the spins within each layer are co-aligned but alternate in orientation between consecutive layers. As previously discussed, FM domain walls in NdBi may in theory host chiral edge states since the gapped topological surface states may bind chiral modes on such boundaries. The FFT of a topography obtained with the Cr tip on such a surface is characterized by a noticeable lack of AFM peaks (Fig 3a inset). This essentially means that the tunneling matrix element is the same for every

Nd atom on the surface with a spin-polarized tip, which indicates uniform magnetism which is as expected for a FM region. Fourier transforms of dI/dV maps on this surface are uniformly featureless (Supplementary Figure S6). The lack of both AFM ordering and QPI are consistent with DFT calculations for the FM surfaces which do not exhibit the surface states or Fermi arcs that were the main contributors to our QPI signal on the AFM surface [27]. We use these features as a fingerprint of FM surfaces.

To experimentally verify the presence of 1D edge states, we examine a step edge (Fig. 3b) as was done in a recent study of MnBi$_2$Te$_4$ [29]. This terrace is characterized by lack of AFM peaks (inset of Fig. 3a) and QPI signal (Supplementary Figure S6) which we have already identified as signatures of the FM surfaces. Odd step edges on the FM domains (Fig.4e) separate terraces of opposite spin and therefore act as magnetic domain walls. The step edge height shown in Fig. 3b is approximately 320 pm, corresponding to a single layer of atoms. This step thus separates two layers with opposite spins and effectively acts as a magnetic domain wall.

Fig. 3a shows dI/dV spectra measured on the terrace and near the step edge. The dI/dV spectrum at the step edge shows that the density of states is larger than the terrace for a wide range of energies above the Fermi energy ($E_F$). This enhanced density of states is clearly visible in dI/dV maps measured of this area (Fig. 3c) which reveal a ~15 Å wide region along the edge with enhanced conductance. As further elaborated on in the discussion section, the fact that the density of states enhancement occurs at energies above $E_F$, is expected according to previous DFT calculations [27].

The existence of the edge state on the step edges of the FM surface is correlated with the onset of the magnetic phase. We measured spectra on the FM surface at different temperatures, below and above $T_{Néel}$, as shown in Supplementary Figure S7. In the paramagnetic phase, the increase in density of states at positive energies is absent, and in fact a monotonic decrease is observed for energies above ~70 meV, with the same spectra seen at different points along a line perpendicular to the edge, shown in Supplementary Figure S8. The measurements show that the edge state exists only below the $T_{Néel}$ and disappears above it along with the magnetism.

As an additional confirmation we examine step edges on the AFM surface. The FM surface, with staggered magnetization on the surface, results in gapped surface Dirac cones, and ultimately in the formation of the chiral edge state. Following this argument, the edge state should not exist on the step edge of an AFM-surface [1]. We identify and measure a suitable AFM step edge in the same area where the FFT reveals strong AFM peaks, as shown in Fig. 3e, f. The dI/dV map on an AFM surface step taken under the same conditions as the FM terrace (Fig. 3g, h), reveals that the edge shows no enhancement and in fact shows a decrease in conductance. This is also reflected in spectra obtained over a broader energy range as illustrated in Supplementary Figure S9.

**Discussion**

Our studies establish the correlation between the surface type and boundary states in NdBi. For AFM surfaces, the presence of electron like surface states and Fermi-arcs generates a strong QPI signal with the same dispersion observed by ARPES. This QPI disappears above $T_{Néel}$ which correlates it with the magnetic ground state and shows that it does not arise from the 3D bands. On the FM surfaces we observe clear 1D edge modes spanning a large energy range. While 1D channels of increased conductance on step edges can also arise from more trivial factors, such as dangling bonds or strong scattering from the step edge itself [37], the absence of 1D edge states on the AFM step edges is strong evidence against such dangling bond states, since the AFM surface step edges and FM surface step edges have identical crystal structures, and differ only in the spin structure. We note that the data on the FM and AFM surfaces was acquired with the same Cr tip, emphasizing the correlation between the FM surface, in the magnetic phase, and the appearance of the edge state.

We now discuss the fact that the edge state appears most clearly at energies above $E_F$ which can be attributed to the increased bulk density of states below $E_F$. To understand this, let's consider the effects of the band folding in the magnetic phase caused by increasing the unit cell size in the z direction. As a result, of the band folding, the band inversion along the $\Gamma$-Z and $\Gamma$-M lines shifts upward in energy. The band inversion along the $\Gamma$-M line spans from negative to positive energies, while the $\Gamma$-Z band inversion is completely above the Fermi level (see Supplementary Figure S10). The magnetization on the FM surface ultimately produces a gapped Dirac cone with a gap of several hundred meV as seen in surface spectral function calculations [26,27] and illustrated in Figure 4. This suggests that the edge state should appear both below and above $E_F$. However, as can be clearly seen from the dI/dV spectra in Figure 3a, e, the density of states rises sharply below $E_F$ which obscures the edge state below $E_F$, Fig. 3d. This scenario is reinforced by the fact that measurements taken on the AFM surface show no clear changes at -50 meV, where the surface state starts dispersing indicating that that signal is also suppressed by the large bulk density of states.

Finally, we note that in addition to the surface states and Fermi arcs, three massless Dirac surface cones exist in both the AFM and paramagnetic phases—one at the Brillouin zone center ($\Gamma$) and two at the M points [25-27]. However, no QPI signal associated with the surface Dirac cones is detected in any QPI measurements, including those performed as a function of temperature. The lack of a QPI signature from the Dirac cones is attributable to their topological nature (spin-momentum locking) which prohibits backscattering. This observation suggests that the other surface states associated with the AFM phase as observed in our QPI data, possess a spin texture that does not forbid backscattering, as previously measured in Weyl semimetals [38,39] and proposed in recent studies on NdBi [27, 30].

The existence of well-localized, chiral step-edge states extends the applicability of the AFM topological insulator picture to metals and materials which host both non-trivial and trivial bands. This is similar to chiral step-edge states observed in the Weyl semimetal phase of $Co_3Sn_2S_2$ where the modes are tightly localized on step edges [24]. Unlike $Co_3Sn_2S_2$ however, where only certain rarely seen terminations show edge modes, rare-earth monopnictides have the advantage that any FM domain should show chiral edge states. Importantly, since the FM surfaces in these materials host gapped Dirac cones and no other surface states or Fermi arcs, the edge states remain isolated and well defined. Our work thus sets the stage for further investigations of 1D chiral edge states, and their applications as building blocks for Majorana quasiparticles in proximitized rare-earth monopnictides and superconductors.

Fig. 1

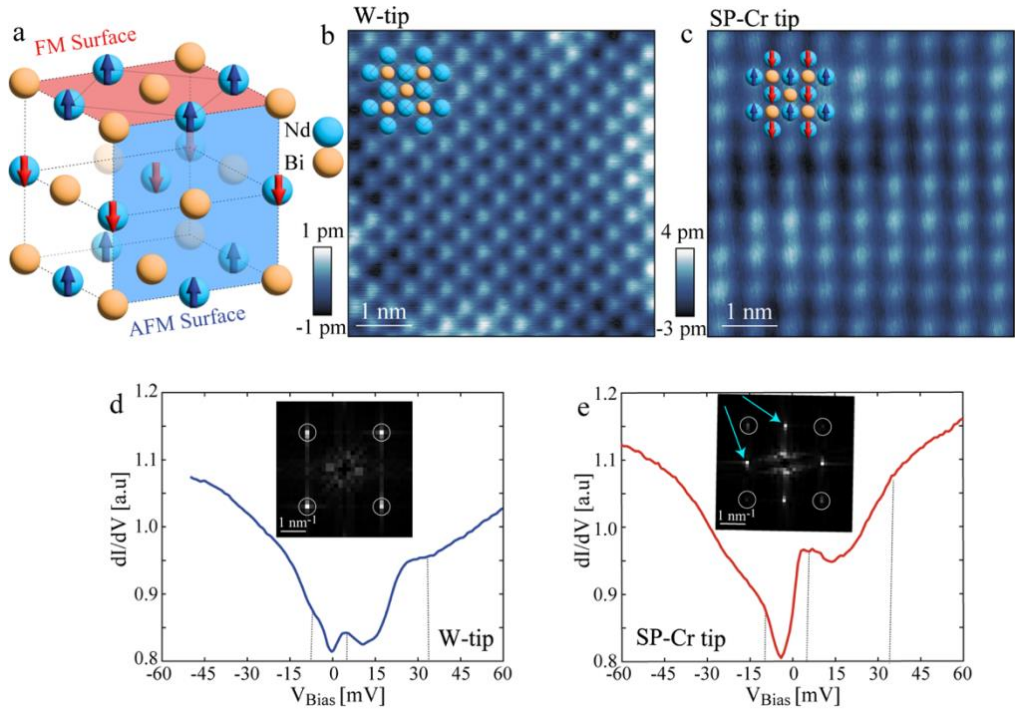

Figure 1 – **NdBi crystal structure and STM topography and spectra**. a Crystal and magnetic structure of NdBi with the Bi atoms in orange and Nd in blue. The crystal structure is rock-salt like, with the spins residing on the Nd atoms. b Topography of the NdBi surface with a W tip, V = 50 meV, I = 700 pA. c Topography of the NdBi surface with a spin polarized Cr tip, V = 25 meV, I = 250 pA. d Averaged spectra taken on the same locations as b with a W tip, $V_{set}$ = 60 meV, $I_{set}$ = 360 pA. inset: FFT of the topography in a with the Bragg peaks (white circles) matching a translation vector with a size of 0.46 mn. e Averaged spectra acquired on the same location as c using a Cr tip, $V_{set}$ = -70 meV, $I_{set}$ = 250 pA. inset: FFT of the topography in c showing the AFM peaks (cyan arrows). The two spectra taken using the W and Cr tip give the same overall features, as shown by the dashed black lines.

Fig. 2

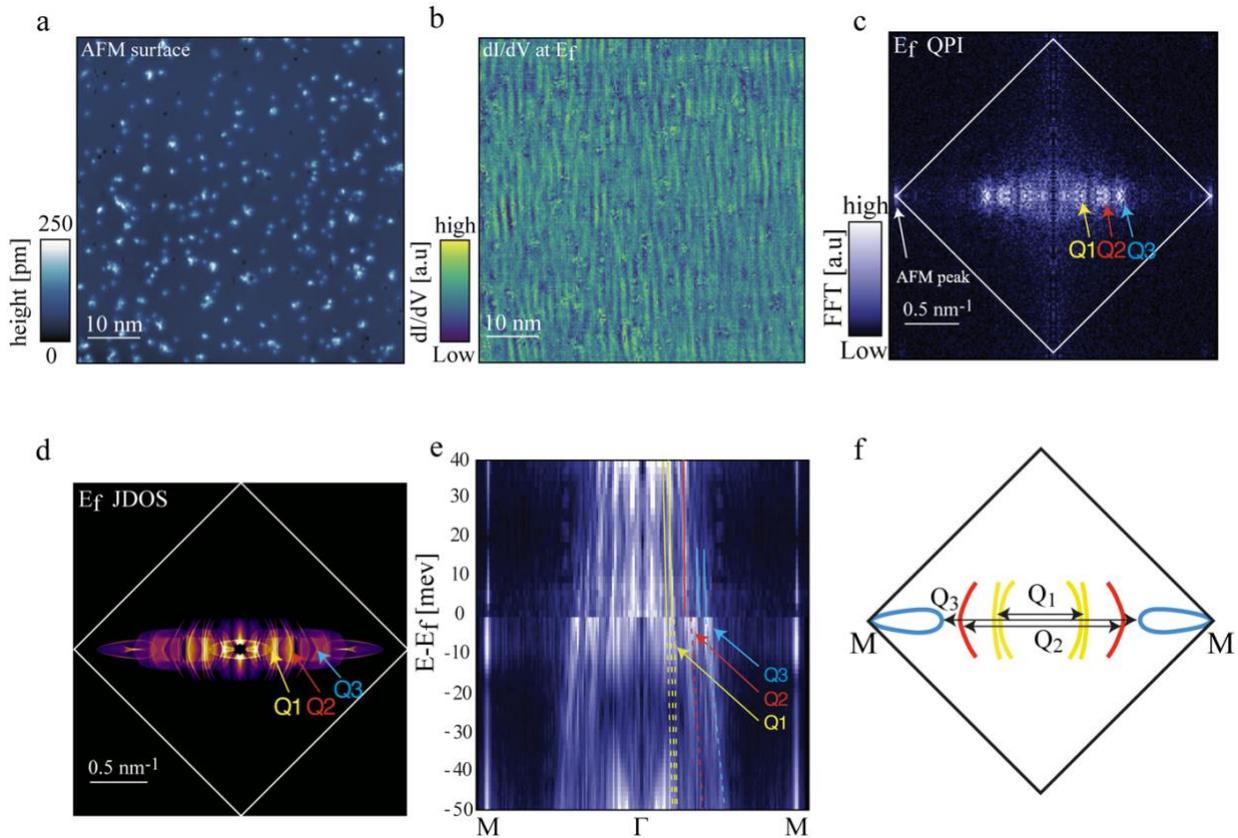

Figure 2 – **QPI data on the AFM surface. a** Topography of an AFM surface with scattered adatoms, where the dI/dV maps were obtained. $V_{set}$ = -50 meV, $I_{set}$ = 160 pA. **b** E=$E_F$ slice of the dI/dV map taken on the region shown in (a), showing the quasiparticles interference patterns with clear C2 symmetry. $I_{set}$ = 160 pA, Vset = -50 meV, $V_{mod}$ = 6 meV, f = 907.5 Hz, **c** FFT of the dI/dV slice in **b**, showing the wave vectors Q1, Q2, Q3 corresponding to scattering of the quasiparticles from the electron like surface state, hole like Fermi arcs and additional surface states around the Γ point. The white square represents the Brillouin zone in the non-magnetic state, with the AFM peak at the M point. **d** Calculated FFT of the QPI patterns of the surface states and Fermi arcs in a single AFM domain obtained by taking the autocorrelation of the Fermi surface shown in **f**. The white square represents the Brillouin zone. **e** QPI dispersion acquired from the map taken in **a** along the Γ-M line, parallel to the Q vectors. The dispersion is visible within a 90 meV energy window. The dashed lines represent the ARPES dispersion from ref [25]. The solid lines are a continuation of the dotted lines with the same slope. The QPI data above and below $E_F$ were obtained in two separate maps. **f** Cartoon of the Fermi surface from ref [25]. The Fermi arcs are shown as yellow and red lines while the electron like surface states are shown using blue line, in the Brillouin zone of the paramagnetic phase with the M point marked. The main scattering vectors are labelled $Q_1$, $Q_2$, $Q_3$, respectively. Q1 represents the scattering within the surface state encircling the Γ point. Q2 corresponds to scattering between the hole like Fermi arcs. Q3 is the scattering between the electron like surface states.

Fig. 3

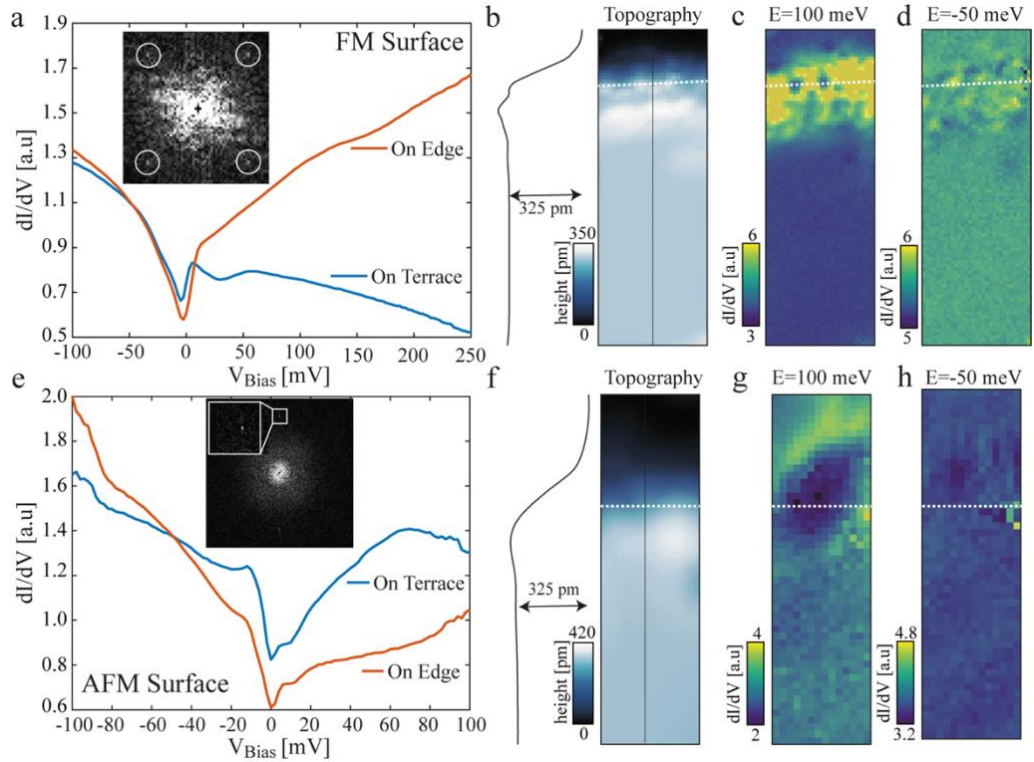

Figure 3 –**Step Edge spectra and maps on a FM nd AFM surface. a** dI/dV spectra acquired on a terrace and on a step edge of a FM surface. An average taken on the edge (orange curve) shows the increased conductance on the step edge relative to the average taken on the terrace (blue curve). Inset: FFT of the topography taken on the terrace showing no AFM peaks. White circles mark the paramagnetic unit cell Bragg peaks. **b** Topography taken in a region with a step edge and the line profile obtained at the solid vertical black line. The height of 320 pm is consistent with a single layer of atoms. V = -100 meV, I = 150 pA. **c** dI/dV map at E= 100 meV on the same area showing the edge state which is localized within 1.5 nm of the edge. **d** dI/dV map at E=- 50 meV. Dashed white lines mark the location of the edge in all panels. **e** Same as **a** for an AFM surface. Inset: FFT of the topography showing AFM peaks. **f, g, h** Topography and dI/dV maps taken on the edge, with a height of a single layer, as in **b**. There is no enhanced density of states on the edge state in either map, emphasizing the absence of the edge state on the AFM surface. V =-100 meV, I = 150 pA. Dashed lines are the same location in all panels, to mark the location of the edge.

Fig. 4

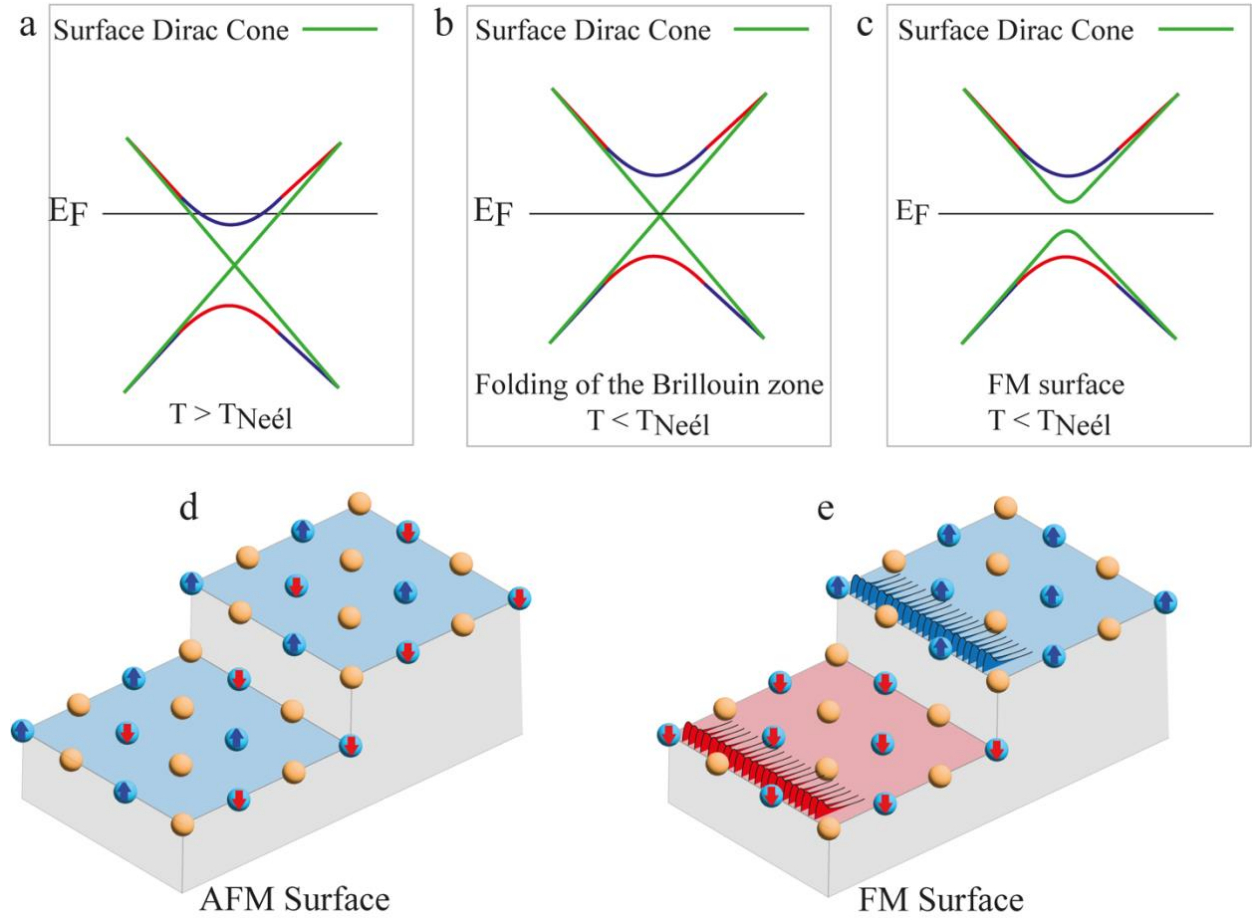

Figure 4 – **Formation of the Chiral Edge State for energies above $E_F$** Cartoon figure depicting the evolution of the band inversion and subsequent surface Dirac cone on the FM surface according to DFT calculations presented in [16]. **a** Above $T_{Néel}$ the material is a strong topological insulator with the band inversion and Dirac point below $E_F$. **b** As the system is cooled into the AFM phase, the doubling of the unit cell in $k_z$ pushes the band inversion to a higher energy (along $\Gamma$-Z and $\Gamma$-M lines). **c** On the FM surface, the Dirac cones are gapped due to the magnetic order and create edge states. **d and e** Step edges of the AFM and FM surfaces, respectively. Odd step edges on the FM surfaces act like domain walls and trap chiral edge states.

## Methods

Single crystals of NdBi were grown out of In flux. The elements with an initial composition of $Nd_6Bi_6In_{94}$ were placed in a 2 ml alumina fritted Canfield Crucible Set [40,41] and sealed in fused silica tube under partial pressure of Argon. The prepared ampules were heated up to 1150° C over 5 hours and held there for 2 hours. This was followed by a slow cooling to 700° C over 120 hours and decanting of the excess flux using a centrifuge [42]. The cubic crystals obtained were stored and handled in a glovebox under Nitrogen atmosphere. STM measurements were performed using a Unisoku STM at an instrument temperature of 1.9K (unless otherwise specified) using chemically etched and annealed tungsten and chromium tips. Spectra were acquired using a standard lockin technique at a frequency of 907 Hz.

## Acknowledgments

STM studies at UIUC were supported by the Air Force Office of Scientific Research (AFOSR) under Grant No. FA9550-23-1-0635. VM acknowledges support from Gordon and Betty More Foundation's EPiQS Initiative through grant GBMF4860 and the Quantum Materials Program at CIFAR where she is a Fellow. A.A acknowledges support from the US National Science Foundation (NSF) Grant Number 2201516 under the Accelnet program of Office of International Science and Engineering (OISE). T. L. H. thanks ARO MURI W911NF2020166 for support. Work done at Ames National Laboratory was supported by the U.S. Department of Energy, Office of Basic Energy Science, Division of Materials Sciences and Engineering. Ames Laboratory is operated for the U.S. Department of Energy by Iowa State University under Contract No. DE-AC02-07CH11358. JAM and HS acknowledge support by the Spanish Research State Agency (PID2020-114071RB-I00, PID2023-150148OB-I00 and CEX2023-001316-M) and the Comunidad de Madrid (TEC-2024/TEC-380).

## Author contributions

A.A. and V.M. conceived the experiments. The single crystals were provided by B.K., J.S., J.A.M and P.C. A.A. obtained the STM data. A.A. and V.M. performed the analysis and R.C. and T.L.H. provided the theoretical input on the interpretation of the data. A.A., V.M. and T.L.H. wrote the paper with input from all authors.

## Competing Interests

The authors declare no competing interests.

# Supplementary information for

# Discovery of a 1D edge mode in a magnetic topological semimetal

Avior Almoalem[1], Rebecca Chan[1,2], Brinda Kuthanazhi[3,4], Juan Scmidt[3,4], Jose A. Moreno[3,5], Hermann Suderow[5], Paul Canfield[3,4], Taylor L. hughes[1,2] & Vidya Madhavan[1]

1. Department of Physics and Materials Research Laboratory, Grainger College of Engineering, University of Illinois at Urbana-Champaign, Urbana, IL, USA
2. Anthony J. Leggett Institute for Condensed Matter Theory, University of Illinois, Urbana, IL, USA
3. Ames Laboratory, Ames, Iowa, USA
4. Department of Physics and Astronomy, Iowa State University, Ames, IA, USA
5. Laboratorio de Bajas Temperaturas y Altos Campos Magnéticos, Unidad Asociada UAM-CSIC, Departamento de Fisica de la Materia Condensada, Instituto Nicolas Cabrera and IFIMAC, Universidad Autonoma de Madrid, E-28049 Madrid, Spain

**Cr tip characterization:**

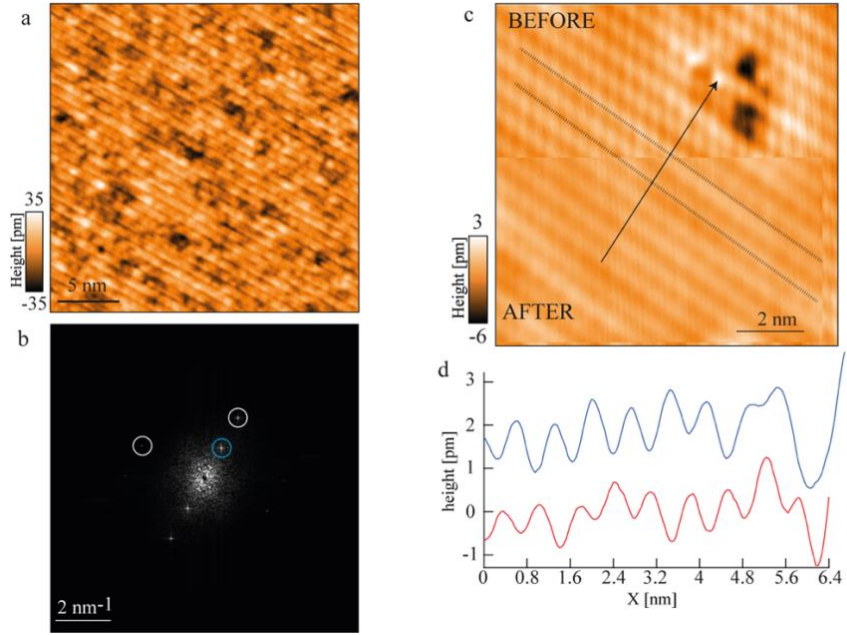

Figure S1 –Characterizing the Cr tip. **a**. Topography scan acquired on a $Fe_{1+x}Te$ crystal with the Cr tip used for all NdBi measurements. V=-30 meV, I =300 pA. **b** FFT of the scan presented in **a** showing both Bragg peaks (white circles) and the AFM stripes peaks (cyan circle) along one of the Bragg peaks and at half of the q vector size, matching the two atomic lattice sites AFM phase in this crystal. **c** Topography scan of the same location before and after changing the tip polarization with a B = -10T magnetic field. The two bright and dark spots are used to align the two scans, before and after polarizing the tip, to show the flipping of the spin. There is one atomic site shift of the atoms as seen by the dashed lines. **d** A line profile taken on the two scans with the same line length, starting at the same point and ending at the same point (the bright defect). The line profiles give the corrugation on the surface before (blue) and after (red) the polarization of the tip. There is a clear shift of the profile. The AFM signal and the tunability of the tip with magnetic field confirm that the tip is indeed spin polarized.

We prepare Cr tips on a $Fe_{1+x}Te$ sample following established procedures [32,33], Supplementary Figure S1. The observed difference in apparent atomic heights, using a Cr tip as opposed to a W tip, is attributed to the spin-selective tunneling process occurring with the Cr tip. A similar effect exists in $Fe_{1+x}Te$, where the AFM state of the Fe lattice is resolved despite the exposed surface being the Te layer, Supplementary Figure S1.

To confirm the spin polarization nature of the Cr tip, we measure the same location at B=0T, once after polarizing the tip at B=4T and then after polarizing the tip at B= -10T. A shift in atom locations is measured, due to the flipping of the spin at the apex of the tip. At first, the tunneling process is mainly into the "up" spins, thus we measure a specific spin orientation. After polarizing the spin at the apex of the tip in the opposite direction, the tunneling is into the "down" spin atoms, thus shifting the scan.

## AFM and FM surfaces additional data:

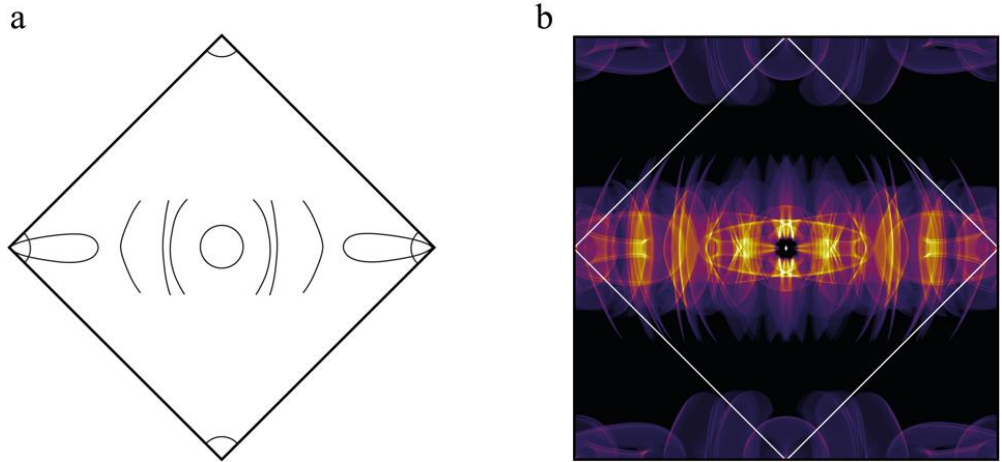

Figure S2 – JDOS simulation of the QPI including the Dirac cones. **a** Cartoon depicting the electron and hole like surface states at $E_F$, including the Dirac cones at the $\Gamma$ and M points of the Brillouin zone. **b** Calculated FFT of the QPI patterns for the surface Dirac cones, electron like surface states and Fermi arcs in a single AFM domain. White square represents the Brillouin zone. Due to the additional Dirac cones at the M point a replica of the original C2 symmetric scattering vectors appear, which is missing in our data.

## QPI maps Raw data:

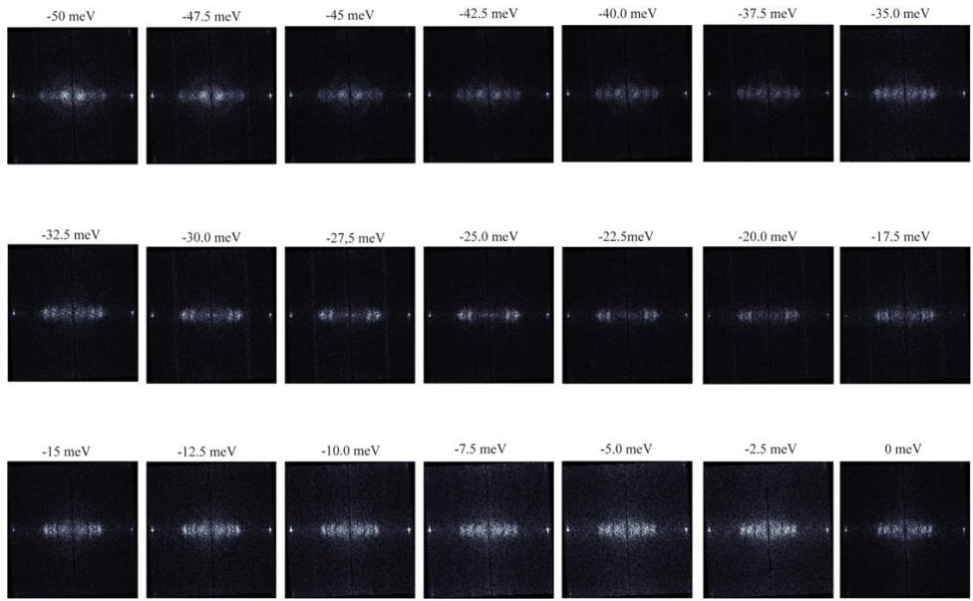

Figure S3 – Additional non-symmetrized FFT data of the dI/dV map acquired on an AFM surface, showing the C2 symmetric data, with no QPI signature in the perpendicular direction to the Q1, Q2, and Q3 vectors as defined in the main text.

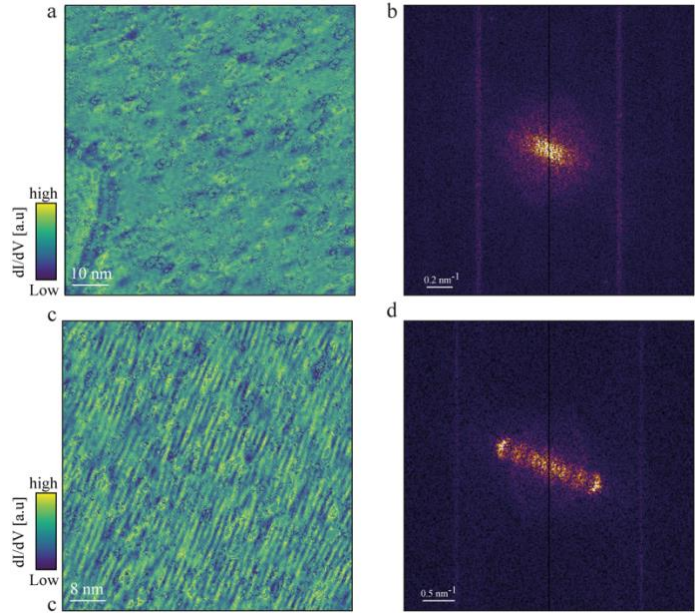

Figure S4 – Conductance maps at different energies, below and at the surface states energies. **a** E=-120 meV slice from a dI/dV map taken with a W tip showing no QPI patterns, and no wave like signature, $I_{set}$ = 450 pA, $V_{set}$ = -120 meV, $V_{mod}$ = 6 meV, f = 907.5 Hz. **b** FFT of the dI/dV map at **a**, with the QPI missing. **c** E= -40 meV slice from a dI/dV map taken with a W tip and the same location as **a** showing the QPI patterns with the wave like signature, $I_{set}$ = 240 pA, $V_{set}$ = -40 meV, $V_{mod}$ = 3 meV, f = 907.5 Hz. **d** FFT of the dI/dV map at **c** showing a clear C2 symmetric QPI signal.

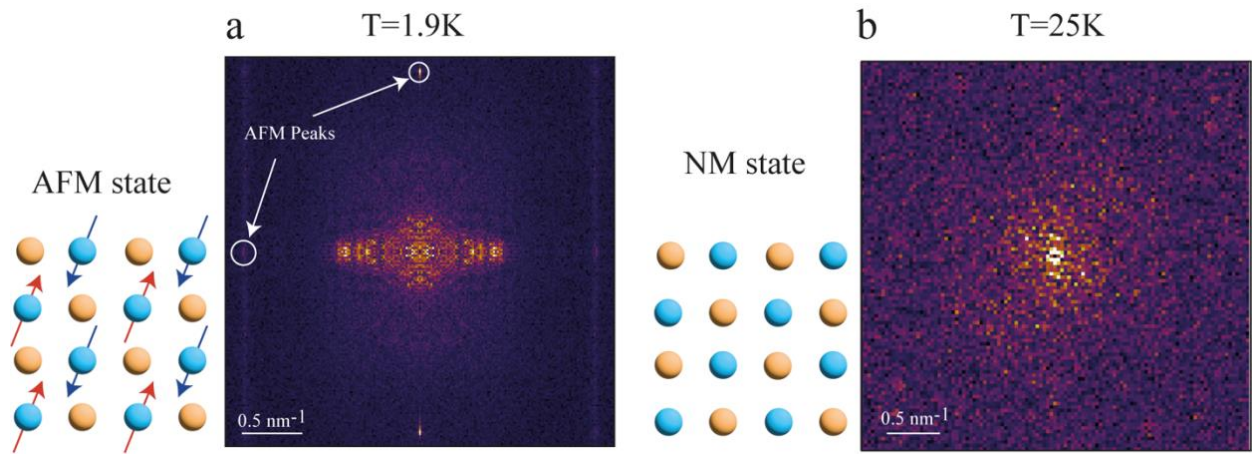

Figure S5 – Evolution of the C2 symmetric QPI with temperature at $E_F$. **a** FFT of the slice at $E_f$ of a conductance map taken at T=1.9K < $T_{Néel}$. AFM peaks are visible at q corresponding to an AFM vector of 0.64 nm. The QPI is along one of the AFM peaks, meaning along the Γ-M line. small panel: cartoon depicting the AFM surface, with the spin and AFM vector in plane. **b** Same as **a** at T=25K > $T_{Néel}$, and at the same location, with the AFM peaks missing and no QPI. small panel: cartoon depicting the non-magnetic surface. The QPI are clearly absent in this map, due to the disappearance of the surface states above Néel temperature. $V_{set}$ = -40 meV, $I_{set}$ = 130 pA, $V_{mod}$ = 6 meV, f = 907.5 Hz.

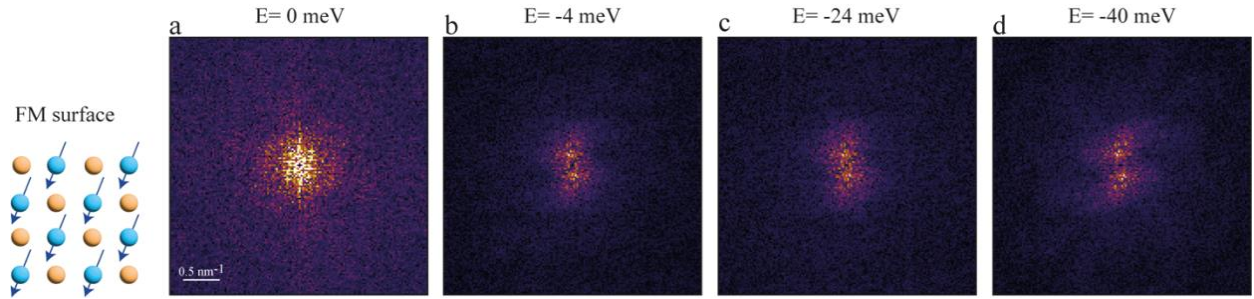

Figure S6 - FFT of conductance maps of different energies along the surface states dispersion taken on a FM surface at T=1.9K<$T_{Néel}$. The AFM peaks are missing, together with the QPI, in **a** 0 meV, **b** -4 meV, **c** -24 meV and **d** -40 meV, as opposed to the corresponding maps taken on the AFM surfaces (supplementary Figure S2). QPI signal is absent due to the disappearance of the surface states as explained in the main text. $V_{set}$ = -40 meV, $I_{set}$ = 200 pA, $V_{mod}$ = 5 meV, f = 907.5 Hz.

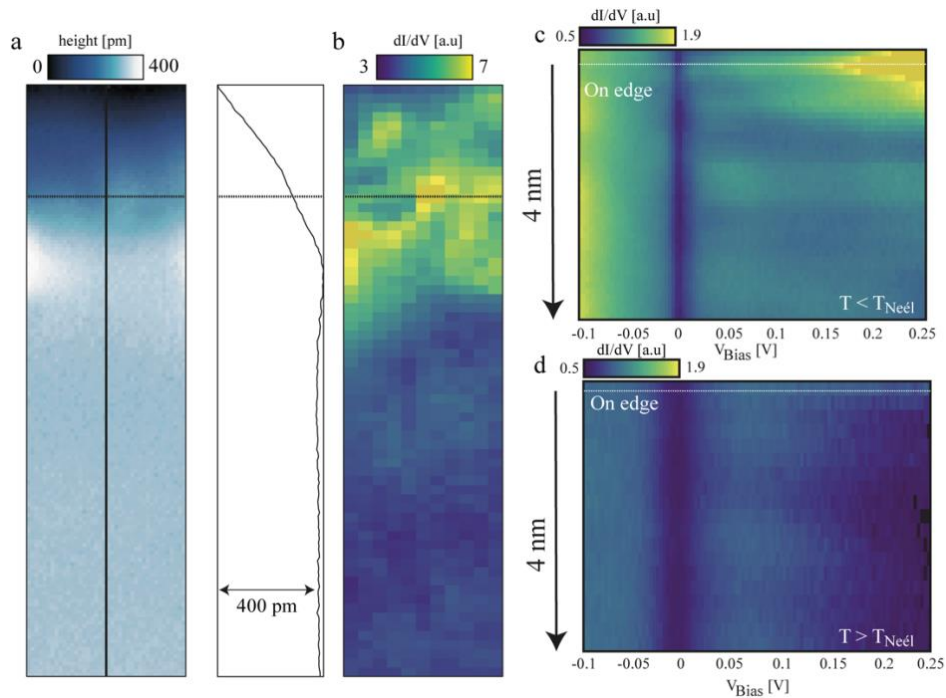

Figure S7 – Additional FM edge state measurements. **a** Topographic image of a FM surface step edge. The total height of the step is about 4.7nm, about 15 times higher than a single unit cell. Map was taken down to 400 pm. V = -150 meV, I = 50 pA. Horizontal line represents the location of the edge. Solid vertical line shows the location of the profile depicted in the panel to the right. Panel: Height profile of the step edge. **b** dI/dV map at E=200meV, showing the 1D chiral-edge state taken at the same location as in **a**. **c** dI/dV spectra acquired along the line shown in **a** below $T_{Néel}$. The 1D channel is shown as an increase in conductance, localized on the edge. **d** Same as **c** above $T_{Néel}$. The enhanced conductance is gone, and the spectra is similar on and off the edge. For **c** and **d** $V_{set}$ = -100 meV, $I_{set}$ = 150 pA.

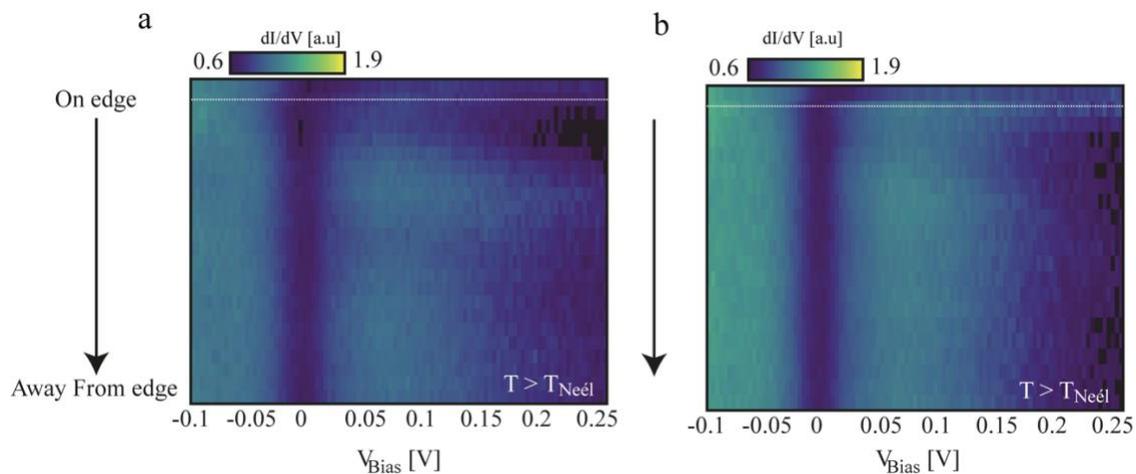

Figure S8 –Absence of an edge state in the paramagnetic phase. **a** dI/dV spectra acquired along the line perpendicular to the edge on a FM surface. The spectra are acquired above $T_{Néel}$. The enhanced conductance is gone on the edge and the spectra is similar on and off the edge. **b** same as **a** in another location on the edge. In both cases the edge is the same edge as in Figure S8. $V_{set}$ = -100 meV, $I_{set}$ = 150 pA.

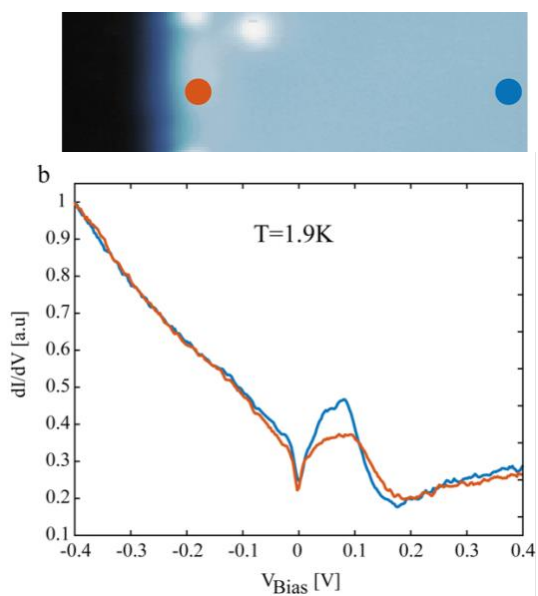

Figure S9– Absence of an edge state on the AFM surface. **a** Topographic image of a single unit cell step edge from an AFM surface to an AFM surface. white dots are adatoms on the surface. inset: FFT showing the AFM peaks taken on the same layer adjacent to the edge. **b** dI/dV spectra locally measured in a region on the step edge (orange) and a few nm away from it (blue). The spectrum on the edge (orange) shows no enhancement of density of states that would be expected for an edge mode, and infact shows a small suppression $V_{set}$ = -400 meV, $I_{set}$ = 500 pA.

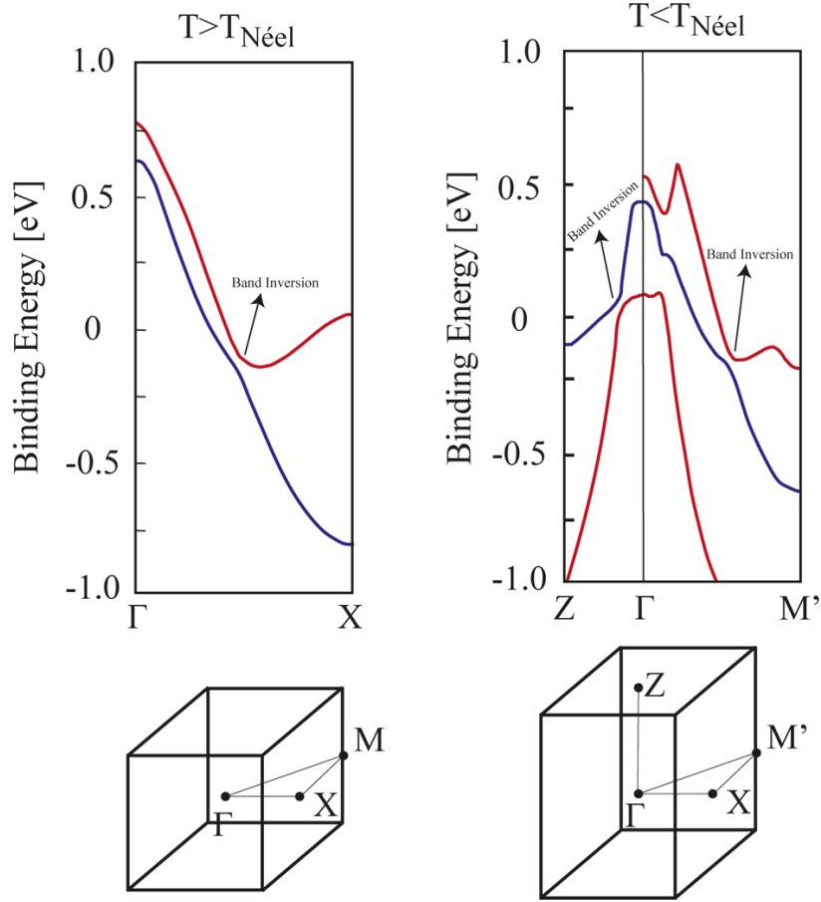

Figure S10 – Band dispersion of the bands around the Fermi level above (left) and below (right) $T_{Néel}$. The dispersion is for the FM surface in the AFM phase showing the band inversion gap is below and above $E_F$ for the Γ-M' line and completely above $E_F$ along Γ-Z. In the paramagnetic phase the band inversion is completely bellow the Fermi level. Data is taken from ref. 16, using the same notations for the high symmetry lines. Lower part shows the Brillouin zones in each phase, note that due to rotation of the unit cell the new M' direction in the AFM phase is along the X direction of the paramagnetic phase.